\documentclass[twoside,12pt]{article}
\usepackage{CJK}
\usepackage{indentfirst}
\usepackage{bm}
\usepackage[dvipsnames]{xcolor}
\usepackage{pdfpages}
\usepackage[dvipsnames]{xcolor}
\usepackage{pdfpages}
\usepackage{epsfig}
\usepackage{hyperref}
\usepackage{amsmath}
\usepackage{bbm}
\usepackage[numbers,sort&compress]{natbib}
\newcommand{\upcite}[1]{\textsuperscript{\cite{#1}}}

\begin{document}

		\begin{center}
			\LARGE\bf The Wiener Path Integral Interpretation of the 3:1 Combat Rule
		\end{center}
		
		\footnotetext{\hspace*{-.45cm}\footnotesize $^\dag$E-mail: zhongm@nudt.edu.cn}

		\begin{center}
			\rm Wei Liang and Ming Zhong$^{\dagger}$
		\end{center}
		
		\begin{center}
			\begin{footnotesize} \sl
				College of Science, National University of Defense Technology, Changsha 410073, People's Republic of China \\
			\end{footnotesize}
		\end{center}
		
		\vspace*{2mm}
		
		\begin{center}
			\begin{minipage}{15.5cm}
				\parindent 20pt\footnotesize 
				The Wiener path integral framework is proposed to model military combat dynamics by incorporating the neglected stochastic effects to the Lanchester's square law.	This framework is applied to evaluate the empirical 3:1 combat rule, which posits that an attacker requires a threefold force superiority to achieve victory. Specifically, the attacker's winning probability is computed utilizing a semi-analytical Rayleigh-Ritz method. Numerical results demonstrate that the validity of the rule critically depends on specific parameter regimes, primarily contingent upon the relative combat effectiveness ratio between the opposing forces and the tolerance for attrition. This work establishes a physics-informed theoretical bridge between statistical mechanics and military operations research for analyzing uncertain combat systems.
			\end{minipage}
		\end{center}
		
		\begin{center}
			\begin{minipage}{15.5cm}
				\begin{minipage}[t]{2.3cm}{\bf Keywords:}\end{minipage}
				\begin{minipage}[t]{13.1cm}	\footnotesize
Wiener path integral; Stochastic dynamics; Transition probability;\\ Military conflict system; Lanchester's model; 3:1 combat rule; Brown motion.
				\end{minipage}\par\vglue8pt
				
			\end{minipage}
		\end{center}
		
		\section{Introduction}
		Stochastic conflict systems constitute a cutting-edge area in modern statistical physics, characterized by emergent collective dynamics arising from competing interactions under uncertainty. The extensive modeling of these systems encompasses a wide range of physical domains, which include but are not limited to predator-prey ecosystems, chemical reaction networks, armed conflict processes, and the statistical mechanics of social impact\upcite{Chen2025,Schmiedl2007,Lee2020,Goto2024,Ispolatov1996,Lewenstein1992,Valenti2016}. Recent progress in path integral method\upcite{Weber2017} has facilitated the first-principles derivation of switching probabilities in such conflict systems, uncovering universal scaling laws in the vicinity of critical points\upcite{Munoz2018}. This physical paradigm offers a natural framework for combat dynamics to which the Wiener path integral method will be employed in this study.
		
		Military combat systems constitute a prototypical class of driven-dissipative conflict systems where stochasticity fundamentally governs attrition dynamics. The Lanchester's models, formalized in 1916\upcite{Lanchester1916}, provide a mechanical deterministic description of force decay via coupled first-order ordinary differential equations. They have been applied extensively in analyzing competitive resource-depletion processes across multiple domains such as combat simulation, market competition, enterprise management, financial gaming, virus transmission, species evolution, and network attack and defense\upcite{Kress2012,Campbell1986,Jorgensen2015,Adams2003,Johnson2015,Clifton2020,Song2021,Taylor1974,Kress2018}. While establishing quantitative foundations for operations research, these classical formulations neglect the intrinsic fluctuations which is an essential physical characteristic of real combat\upcite{Kao2014}. Existing stochastic extensions, primarily Markovian state-transition models\upcite{Koopman1970,Kress1999,McNaught1999,Morse1950,Karr1983} and stochastic differential equations\upcite{Amacher1986,Zadorozhny2021}, suffer from limited applicability, poor generalizability, and the lack of a systematic mathematical foundation. They typically fail to capture the physics-informed features because of the limitations of the discrete state-space and the memoryless assumptions.
		
		The Wiener path integral framework\upcite{Chaichian2001}, grounded in statistical physics\upcite{Mauri1986,Leporini2007,Chaichian2009}, may provide a mathematically rigorous framework to overcome these limitations. In this work, we address this purpose and establish a novel nonlinear Wiener path integral formalism to model stochastic combat dynamics, building upon Lanchester's square law. This approach conceptualizes a combat system as a complex physical system of two interacting Brownian particles. It enables the computation of the probability distribution over all possible final states transitioning from a given initial state. This computed distribution is then applied to interpret the empirical 3:1 combat rule.
		
		The so-called ``3:1 combat rule'' persists as a controversial topic in military analysis. As per reference \upcite{Mearsheimer1989}, it is formulated as an inductive rule stating that an attacker must possess an advantage of more than 3:1 on each main axis to achieve victory, which has garnered substantial attention. Nevertheless, it has been challenged on multiple fronts. There are ambiguities in the definition of proportional units, a dearth of empirical evidence, and a lack of methodological rigor in its derivation\upcite{Epstein1988,Epstein1989}. Similarly, due to the absence of reliable combat data for validation, it has been regarded as a rudimentary heuristic principle\upcite{Dupuy1989}. In response to these issues, several analytical methodologies have been adopted, including adaptive dynamic mode modeling\upcite{Epstein1988}, and stochastic extensions of the Lanchester equations\upcite{Kress1999}. Notably, the stochastic model presented in the reference \upcite{Kress1999} integrated probabilistic factors, thus overcoming the limitations inherent in earlier deterministic models. Given these ongoing controversies, there is an evident necessity to reevaluate the 3:1 combat rule using other mathematical frameworks. 
		
        In the next section, we will briefly review the Lanchester's model. A concise introduction to the Wiener path integral is provided in Section 3 to characterize interacting Brownian particles. In Section 4, the Lanchester's square law is naturally incorporated into the Wiener path integral formalism to consider the inherent battlefield fluctuations. Subsequently, in Section 5, a Wiener path integral analysis of the 3:1 combat rule is presented. We determine the temporal evolution of the attacker's winning probability under various parameters that govern the strength of the attacker and the defender. It is found that the effectiveness of the rule, regardless of whether it benefits the attacker or the defender, is critically contingent upon the relative combat efficiencies of the opposing forces and the tolerance for attrition, which is in line with the conclusion derived from the other stochastic combat model\upcite{Kress1999}. Section 6 is for a short summary, and two appendixes are for the derivation of the diffusion matrix of Lanchester's square law and the Rayleigh-Ritz method to solve the Wiener path integral.
		
		\section{The Lanchester's model}
		The Lanchester's model constitutes an empirical formulation describing the attrition of opposing forces. This section presents a concise introduction to the classical Lanchester's model (for more details please refer to the reference \upcite{MacKay2006}). Considering an isolated system comprising red and blue armies, the Lanchester's model can be expressed as
		\begin{equation}
			\label{LanQ}
			\left\{
			\begin{aligned}
				\frac{\mathrm{d}q_\mathrm{r}(t)}{\mathrm{d}t} &= F_\mathrm{r}(q_\mathrm{r}(t),q_\mathrm{b}(t))\\
				\frac{\mathrm{d}q_\mathrm{b}(t)}{\mathrm{d}t} &= F_\mathrm{b}(q_\mathrm{r}(t),q_\mathrm{b}(t))
			\end{aligned}
			\right.,
		\end{equation}
		where $q_\mathrm{r}(t)$ and $q_\mathrm{b}(t)$ denote the troop strengths of the Red and Blue forces at time $t$, respectively. The function $F_\mathrm{r,b}(q_\mathrm{r}(t), q_\mathrm{b}(t))$ acts as the driving force governing the dynamics of both $q_\mathrm{r}(t)$ and $q_\mathrm{b}(t)$, where distinct functional forms correspond to different Lanchester equations. From this perspective, the Lanchester equations share a fundamental equivalence with Newton's second law of motion in physics, i.e. $\frac{\mathrm{d}\boldsymbol{p}}{\mathrm{d}t}=\boldsymbol{F}$, where the applied force equals the time derivative of momentum. This establishes the Lanchester equations as the Newton's second law for combat systems. 
		
		These troop deployment patterns are denoted as the Lanchester First Linear Law, Second Linear Law, and square Law, applicable respectively to ancient cold-weapon combat, modern indirect-fire targeting operations, and direct-fire engagements. The ``forces'' acting upon the red and blue sides under First Linear Law and Second Linear Law are defined as
		\begin{equation}
			\label{lf1}
			F_\mathrm{r}(q_\mathrm{r}(t),q_\mathrm{b}(t))=-\rho_\mathrm{b},\qquad F_\mathrm{b}(q_\mathrm{r}(t),q_\mathrm{b}(t))=-\rho_\mathrm{r},
		\end{equation}
		and
		\begin{equation}
			\label{lf2}
			F_\mathrm{r}(q_\mathrm{r}(t),q_\mathrm{b}(t))=-\rho_\mathrm{b} q_\mathrm{r}(t)q_\mathrm{b}(t),\qquad F_\mathrm{b}(q_\mathrm{r}(t),q_\mathrm{b}(t))=-\rho_\mathrm{r} q_\mathrm{r}(t)q_\mathrm{b}(t),
		\end{equation}
		where $\rho_\mathrm{r}$ and $\rho_\mathrm{b}$ are time-invariant constants independent of $q_\mathrm{r}(t)$ and $q_\mathrm{b}(t)$, reflecting the combat effectiveness coefficients of the red and blue sides. The ratio of the ``force'' magnitudes acting on both sides remains constant
		\begin{equation}
			\frac{F_\mathrm{r}(q_\mathrm{r}(t),q_\mathrm{b}(t))}{F_\mathrm{b}(q_\mathrm{r}(t),q_\mathrm{b}(t))}=\frac{\rho_\mathrm{b}}{\rho_\mathrm{r}}.
		\end{equation}
		Given the ``force'' forms Eq.\eqref{lf1} and Eq.\eqref{lf2}, the solution to Eq.\eqref{LanQ} satisfies a linear law
		\begin{equation}
			\label{linlaw}
			\rho_\mathrm{r}q_\mathrm{r}(t)-\rho_\mathrm{b}q_\mathrm{b}(t) = \text{constant}.
		\end{equation}
		
		The ``force'' exerted on both sides under the square law is defined as
		\begin{equation}
			\label{sql}
			F_\mathrm{r}(q_\mathrm{r}(t),q_\mathrm{b}(t))=-\rho_\mathrm{b} q_\mathrm{b}(t),\qquad F_\mathrm{b}(q_\mathrm{r}(t),q_\mathrm{b}(t))=-\rho_\mathrm{r} q_\mathrm{r}(t).
		\end{equation}
		The solution to Eq.\eqref{LanQ} yields the temporal evolution of red and blue force strengths
		\begin{equation}
			\label{lsls}
			\left\{
			\begin{aligned}
				q_\mathrm{r}(t) &= q_\mathrm{r}(0) \cosh (\sqrt{\rho_\mathrm{r} \rho_\mathrm{b}}\, t) - \sqrt{\frac{\rho_\mathrm{b}}{\rho_\mathrm{r}}} q_\mathrm{b}(0) \sinh (\sqrt{\rho_\mathrm{r} \rho_\mathrm{b}}\, t) \\
				q_\mathrm{b}(t) &= q_\mathrm{b}(0) \cosh (\sqrt{\rho_\mathrm{r} \rho_\mathrm{b}}\, t) - \sqrt{\frac{\rho_\mathrm{r}}{\rho_\mathrm{b}}} q_\mathrm{r}(0) \sinh (\sqrt{\rho_\mathrm{r} \rho_\mathrm{b}}\, t)
			\end{aligned}
			\right.,
		\end{equation}
		where $q_\mathrm{r}(0)$ and $q_\mathrm{b}(0)$ represent the initial force strengths of the red and blue sides. This solution satisfies the following square law
		\begin{equation}
			\label{sqlaw}
			\rho_\mathrm{r}q^2_\mathrm{r}(t)-\rho_\mathrm{b}q^2_\mathrm{b}(t) = \text{constant}.
		\end{equation}
		
		The Lanchester's model Eq.\eqref{LanQ} describes a simplified combat scenario between two isolated systems without reinforcements. The left-hand side quantifies the rate of attrition for one force per unit time, while the right-hand side represents the attrition inflicted by the opposing force. This model, comprising the force differential equation \eqref{LanQ} and its initial conditions, exemplifies mechanistic determinism. Given the combat effectiveness coefficients $\rho$,$\beta$ and initial force strength $q(0)$, the force strength $q(t)$ at any time $t$ is uniquely determined by solving the equation, as shown in solution Eq.\eqref{lsls}. Furthermore, it embodies an extreme mechanistic deterministic proposition of ``the winner always wins'' arising from inherent conservation laws expressed in Eq.\eqref{linlaw} and Eq.\eqref{sqlaw}. Here, $Q_1=\rho_\mathrm{r}q_\mathrm{r}-\rho_\mathrm{b}q_\mathrm{b}$ and $Q_2=\rho_\mathrm{r}q^2_\mathrm{r}-\rho_\mathrm{b}q^2_\mathrm{b}$ constitute invariant of the system. Provided the combat effectiveness coefficients remain constant during an engagement, if initial force deployments satisfy $Q_{1,2}>0$, the red force maintains a consistent advantage throughout the conflict.
		
		Certain analytical outcomes derived from the Lanchester equations align with established principles of warfare, thereby constituting a fundamental tool of force attrition and conflict evolution. This model represents a prevalent mathematical framework for combat simulation research and has been applied to analyze data from historical battles including Iwo Jima, Kursk, and the Ardennes\upcite{Engel1954,Bracken1995,Hartley1995,Fricker1998,Lucas2004(2),Johnson2011}, as well as the ongoing conflict in Ukraine\upcite{Chung2024}. 
			
		In practice, however, such deterministic behavior stands in stark contrast with the frequently unpredictable outcomes observed in actual warfare. Factors including weather, judgment errors, personnel performance, equipment failures, intelligence acquisition, unforeseen events, and luck can profoundly influence the course and outcome of military engagements. These stochastic effects render combat an extremely complex system characterized by pervasive uncertainty, resulting in divergence between model simulations of specific historical engagements and documented outcomes, and yielding unsatisfactory results\upcite{Fricker1998,Lucas2004(2)}.
		
		To characterize and quantify the dynamic impact of such factors on combat processes, the stochastic Lanchester equation was proposed. By incorporating stochastic processes and probabilistic structures, it aims to provide a mathematically tractable framework for modeling uncertainty in warfare. For one thing, models grounded in Markov process theory introduce the Markov property into the Lanchester's framework\upcite{Koopman1970,Kress1999,McNaught1999,Morse1950,Karr1983}, enabling derivation of the system's probability distribution through state transition probabilities. This facilitates computation of key metrics, including the force distribution at combat termination and the expected engagement duration. For another, analytical formulations based on stochastic differential equations primarily employ tools from stochastic analysis\upcite{Amacher1986,Zadorozhny2021}, such as the Itô integral. These formulations introduce stochastic noise terms into the original differential equations or directly model force dynamics as Itô stochastic differential equations. Subsequently, theories of stochastic differential equations and the Itô lemma are applied to solve for the probability distribution of combat outcomes.
		
		Although the theory of stochastic Lanchester equations has extended the original model and mitigated certain inherent limitations, it still suffers from several shortcomings, including restricted applicability, limited generalizability, and the absence of a systematic mathematical foundation. In the present study, the aforementioned disadvantages will be surmounted through the introduction of the Wiener path integral description.
	
		\section{Introduction to Wiener path integral}
		\label{Wiener}
		The Wiener path integral method, a mathematical formalism for characterizing the stochastic trajectories of classical particles, provides distinct advantages in investigating the dynamical behavior of complex stochastic systems\upcite{Chaichian2001}. By comprehensively integrating over all possible random paths, this technique constructs probability distributions governing system evolution, thereby furnishing a rigorous mathematical framework for analyzing phenomena including Brownian motion and diffusion processes. Its theoretical foundation amalgamates stochastic process theory with functional integral methodology, thus enabling not only the derivation of probabilistic laws dictating system evolution but also the natural incorporation of fluctuations and non-equilibrium characteristics. 
		
		In statistical physics, the Langevin equation of multi-dimensional Brownian particles interacting with each other and moving in the presence of a linear external force reads as
		\begin{equation} \label{Langevin equaion}
			\dot{\boldsymbol{x}}(t)+\boldsymbol{k}\boldsymbol{x}(t) =\bm{\phi}(t).
		\end{equation}
		The n-dimensional vector $\boldsymbol{x}=(x_1,x_2,...,x_n)^T$ is the x-coordinate of n Brownian particles, and $\dot{\boldsymbol{x}}$ denotes its time derivative.The n-dimensional matrix $\boldsymbol{k}=(k_{ij}) _{n\times n}$ refers to the interaction coefficients. The Gaussian white noise vector $\bm{\phi }(t) =(\phi _1,\phi _2,...,\phi _n)^T$ denotes n-dimensional random force characterized by the expected values $\mathrm{E}\left[ \phi _i(t) \right] =0$ and $\mathrm{E}\left[ \phi _i(t) \phi _j(s) \right] =\sigma^2_{ij}\delta_{ij}\delta \left( t-s \right)$. The diagonal matrix ${\boldsymbol{D}}=\text{diag}(\sigma^2_{11},\sigma^2_{22},...,\sigma^2_{nn})$ describes the magnitude of the intensity of diffusive motion. It is typically derived by analyzing the Langevin equation (Eq.\eqref{Langevin equaion}) and statistical regularities embedded in generalized coordinate fluctuations. 
		
		The Langevin equation is the microscopic basis for the Fokker–Planck equation and, in particular, for the diffusion equation
		\begin{equation}\label{Fokker-Planck equation}
			\frac{\partial W}{\partial \tau}+\sum_i\frac{\partial (A_iW)}{\partial x_i}-\sum_i\frac{\partial^2 (B_{ii}W)}{\partial x_i^2}=0,
		\end{equation}
		where $W(\boldsymbol{x}(\tau),\tau|\boldsymbol{x}(0),0)$ is the transition probability density of the Brownian particle moving from the initial spatio-temporal coordinate $(\boldsymbol{x}(0),0)$ to the terminal coordinate $(\boldsymbol{x}( \tau),\tau)$. The vector $\boldsymbol{A} = -\boldsymbol{kx}$ represents the deterministic influence on the Brownian motion. The matrix $\boldsymbol{B}=\frac{1}{2}\boldsymbol{D}$ is associated with the effect of the noise. The equation depicts the temporal evolution of the position probability distribution of Brownian particles. Given the initial condition
		\begin{equation}
			W(x_1(t),x_2(t),t|x_1(0),x_2(0),0)\overset{t\to0}{\longrightarrow} \delta(x_1(t)-x_1(0))\delta(x_2(t)-x_2(0)),
		\end{equation}
		the solution of the above equation can be formulated via the Wiener path integral (for more details, please refer to the chapter 1 of reference \upcite{Chaichian2001})
		\begin{equation} \label{path integral formalism}
			W(\boldsymbol{x}(\tau),\tau|\boldsymbol{x}(0),0) =\mathrm{e}^{\frac{\text{Tr}[\boldsymbol{k}] \tau}{2}}\int_{\mathcal{C}\{\boldsymbol{x}(0),0;\boldsymbol{x}(\tau),\tau\}}{\mathcal{D}}\boldsymbol{x}(t)\mathrm{e}^{-\int_0^{\tau}\mathrm{d}t\left[ \boldsymbol{\dot{x}}+\boldsymbol{k}\boldsymbol{x} \right] ^T\left[ 2\boldsymbol{D} \right] ^{-1}\left[ \boldsymbol{\dot{x}}+\boldsymbol{k}\boldsymbol{x} \right]},
		\end{equation}
		where $\mathcal{D}\boldsymbol{x}(t)$ is the Wiener functional integral measure, and ${\mathcal{C}\{\boldsymbol{x}(0),0;\boldsymbol{x}(\tau),\tau\}}$ is the set of all possible trajectories for the system evolution. The integrand on the exponent $\mathcal{L}(\boldsymbol{\dot{x}},\boldsymbol{x},t)=\left[ \boldsymbol{\dot{x}}+\boldsymbol{k}\boldsymbol{x} \right] ^T\left[ 2\boldsymbol{D} \right] ^{-1}\left[ \boldsymbol{\dot{x}}+\boldsymbol{k}\boldsymbol{x} \right]$ is the Lagrangian function, and $\mathcal{S}=\int_0^{\tau}\mathrm{d}t\mathcal{L}(\boldsymbol{\dot{x}},\boldsymbol{x},t)$ is the action functional. It postulates that the probability density for a particle transitioning from an initial position \(\boldsymbol{x}(0)\) to a final position \(\boldsymbol{x}(t)\) is equivalent to the summation over the contributions from all possible paths connecting these two points. The contribution of each individual path is dictated by its respective probability weight prescribed by the action functional.
		
		Recent developments in Wiener path integral research have focused on solving the integral and extending its application beyond Gaussian noise regimes. The integration of sophisticated analytical and numerical techniques include closed-form estimation, Rayleigh-Ritz approximation, quadratic action functional expansion, free-boundary estimation, and nonlinear filtering\upcite{Petromichelakis2021,Zhang2025,Mavromatis2024,Meimaris2020,Psaros2018,Zhang2023}. Extension to non-Gaussian noise environments, particularly L\'{e}vy noise\upcite{Pirrotta2017,Xu2019,Zan2021,Zan2022,Zan2023,Peng2024(2)} and Poisson noise\upcite{Zan2022(2),Zhang2024,Peng2024,Jia2026}, has significantly expanded the method's applicability. These advances have enabled the method to effectively address complex system characteristics, including high dimensionality, nonlinearity, non-Gaussian noise, and fractional calculus operators, thereby substantially expanding its applicability and modeling capabilities.
		
	The semi-classical approximation constitutes the most established and formally concise approach for solving Wiener path integral. This method operates by fixing terminal system states and incorporating these boundary conditions into the Euler-Lagrange equations to derive analytical approximations. For low-dimensional, non-interacting systems, solutions obtained via this approach exhibit close agreement with the exact analytical solutions. However, when applied to high-dimensional, coupled complex systems, the semi-classical framework frequently encounters limitations: computational complexity scales prohibitively with system dimensionality, often precluding practical implementation or yielding non-closed-form analytical expressions. 
	
	To illustrate a two-dimensional system with given initial state \((x_{1}(0), x_{2}(0), 0)\) and final state \((x_{1}(\tau), x_{2}(\tau), \tau)\), we utilize the semi-classical approximation approach to derive an analytical solution for the path that minimizes the action. The transition probability density \\$W(x_{1}(\tau), x_{2}(\tau), \tau | x_{1}(0), x_{2}(0), 0)$ can then be computed to high precision via an expansion around this minimal-action path. Extremizing the action functional
		\begin{equation*}
			\delta \mathcal{S}=\delta \int_0^{\tau}\mathrm{d}t\mathcal{L}(\boldsymbol{\dot{x}},\boldsymbol{x},t)
		\end{equation*}
	yields the Euler-Lagrange equations 
		\begin{equation*}	
			\left\{ \begin{array}{l}\frac{\partial \mathcal{L}}{\partial x_1}- \frac{\mathrm{d}}{\mathrm{d}t} \frac{\partial \mathcal{L}}{\partial \dot{x_1}}=0
				\\ \frac{\partial \mathcal{L}}{\partial x_2}- \frac{\mathrm{d}}{\mathrm{d}t} \frac{\partial \mathcal{L}}{\partial \dot{x_2}}=0
			\end{array} \right. .
		\end{equation*}
		Solving the equations under the specified initial conditions results in the minimal-action trajectories \(x_1(t)\) and \(x_2(t)\). Substituting these trajectories into the transition probability density gives
		\begin{equation}
			\label{W2d}
			W(x_{1}(\tau), x_{2}(\tau), \tau | x_{1}(0), x_{2}(0), 0)=N \mathrm{e}^{-\int_{0}^{\tau}\mathrm{d}t\mathcal{L}( \dot{x_1}(t),\dot{x_2}(t),x_1(t),x_2(t),t) },
		\end{equation}
		with $N$ the normalization constant. This solution is the leading order of the transition probability expanded around the minimal-action path.

		\section{Wiener path integral formalism of Lanchester's square law}
		To illustrate the application of the path integral approach in military operations analysis, we use the Lanchester's square law (Eq.\eqref{sql}) as a fundamental example and present its corresponding path integral formulation. Such application can be also employed to other combat models. For clarity, the model is reformulated following the convention of Eq.\eqref{Langevin equaion}
		\begin{equation}\label{Lanchester equation}
			\left\{ \begin{array}{l}
				\dot{x_1}=-\beta x_2\\
				\dot{x_2}=-\rho x_1\\
			\end{array} \right. .
		\end{equation}
		The $x_1$ and $x_2$ are strength forces of two opposing sides in a combat, with the coupling coefficients $\rho ,\beta$ denoting their combat efficiencies accordingly.
		
		The combat system depicted by the Lanchester's square law can be effectively compared to a pair of Brownian particles that interact with each other. This analogy regards the two combating troops as a pair of particles, with their ``positions'' defined by their current strength levels. A troop's attrition rate is determined by the size of its opponent, just like a force field that governs the motion of the particles. The inevitable uncertainties in warfare, such as the fog of war or command errors, play a role analogous to the random thermal noise that affects the particles. This conceptual parallel offers a solid basis for reformulating the Lanchester equations into a two-dimensional Langevin equation, a standard framework in physics describing systems subject to both deterministic and random forces. In this formulation, the advanced mathematics of the Wiener path integral may be directly applied to compute the transition probability density $W(x_1(\tau),x_2(\tau),\tau|x_1(0),x_2(0),0)$ of the strength forces. This technique enables us to analyze the entire conflict by calculating the probability of all possible paths that the battle may take, transcending a single, pre-determined outcome. 
		
		Therefore, the Langevin equation and the transition probability density of the Lanchester's square law are thus special cases of Eq.\eqref{Langevin equaion} and Eq.\eqref{path integral formalism} with the matrix 
		\begin{equation*}
			\boldsymbol{k}=\left( \begin{matrix}
				0&		\beta\\
				\rho&		0\\
			\end{matrix} \right).
		\end{equation*}
		
		The stochastic effects inherent in warfare are proportional to operational intensity, particularly manifesting as an intensifying fog of war during large-scale engagements. To achieve this mathematical characterization of combat uncertainty, the stochastic noise must be formulated as a state-dependent multiplicative term, resulting in a diffusion coefficient $\sigma \propto \sqrt{x_1x_2}$. The formal derivation is presented in the Appendix A. This yields the state-dependent nonlinear diffusion matrix
		\begin{equation}{\label{diffusion coefficient matrix}}
			\boldsymbol{D} =\left( \begin{matrix}
				\beta x_1x_2&		0\\
				0&		\rho x_1x_2\\
			\end{matrix} \right).
		\end{equation}
		Then, the Lagrangian function for the Lanchester's square law is
		\begin{equation}{\label{nonlinear Lanchester square law}}
			\mathcal{L}=\frac{1}{2}\frac{(\dot{x_1}+\beta x_2)^2}{\beta x_1x_2}+\frac{1}{2}\frac{(\dot{x_2}+\rho x_1)^2}{\rho x_1x_2}.
		\end{equation}

		As compared to the existing stochastic Lanchester's models, the Wiener path integral formulation adopted in this study demonstrates some advantages for analyzing force attrition dynamics. For one thing, the path integral framework inherently accommodates high-dimensional and nonlinear interactions without reliance on state space discretization, thereby offering superior generalizability. For another thing, it provides enhanced physical intuitiveness, being grounded in the Lagrangian formalism and aligning with principles of statistical physics. This enables direct interpretation of probability evolution through dominant paths and natural derivation of fluctuation properties via generating functionals. And thirdly, by leveraging the systematic mathematical theory of path integral, the method unifies constraint handling within a continuous framework under specific combat scenarios, thereby circumventing ad hoc state-space adjustment strategies commonly required in Markovian discrete-state models.
		
		\section{Wiener path integral analysis of the 3:1 combat rule}
					
		The 3:1 rule stipulates that the attacking force should possess at least triple the combat strength of the defender to achieve victory. This principle, derived from empirical observations, represents more or less a statistical law. Consequently, the Wiener path integral formalism provides an ideal framework for its investigation, specifically enabling the evaluation of the probability distribution given by Eq.\eqref{W2d} using the Lagrangian function defined in Eq.\eqref{nonlinear Lanchester square law}, under the conditions prescribed by this rule.
		
		The $x_1$ and $x_2$ refer to the forces of the attacker and the defender respectively. For clarity, $(x_{10},x_{20})$ is the initial state and $(x_{1t},x_{2t})$ is the state at some later time $t$. Analogously to the reference \upcite{Kress1999} the attacker's winning probability $P_\mathrm{1w}$ is defined to be
		\begin{equation} \label{win probability}
			P_\mathrm{1w} =(1+\frac{f_1x_{10}}{f_2x_{20}}R ) ^{-1},
		\end{equation}
		 as the critical state occurring when the defender reaches its collapse threshold $(1-f_2)x_{20}$ and the attacker remains above its metastability basin. The parameters $f_1=\frac{x_{10}-x_{1t}}{x_{10}}$ and $f_2=\frac{x_{20}-x_{2t}}{x_{20}}$ are attrition rates of the attacker and the defender respectively. In the following analysis, the magnitude of their values signifies the capacity of offensive and defensive units to withstand attrition. The ratio $R$ is  
		\begin{equation} \label{the ratio R}
			R=\frac{\int_{\left( 1-f_2 \right) x_{20}}^{x_{20}}{W\left( \left( 1-f_1 \right) x_{10},x_{2t},t|x_{10},x_{20},0 \right) \mathrm{d}x_{2t}}}{\int_{\left( 1-f_1 \right) x_{10}}^{x_{10}}{W\left( x_{1t},\left( 1-f_2 \right) x_{20},t|x_{10},x_{20},0 \right) \mathrm{d}x_{1t}}}.
		\end{equation}
		
		The ratio $\alpha = \beta / \rho$ quantifies the relative combat effectiveness between the defender and the attacker. The magnitude of $\alpha$ reflects the military capabilities of both belligerents—encompassing tactical proficiency, armaments, and combat resolve—while simultaneously signifying their respective roles as defender and attacker within the conflict. When opposing forces possess comparable military capabilities, the defender typically benefits from advantages in environmental familiarity, physical endurance, and concealment, resulting in superior combat effectiveness relative to the attacker. Following the approach established in the reference \upcite{Kress1999}, we adopt the critical threshold ratio for a draw outcome in the deterministic Lanchester's model as the value for $\alpha$ under such conditions, corresponding to the well-established 3:1 rule. A draw occurs when the invariant quantity derived from Eq.\eqref{sqlaw} satisfies $Q_2 = \rho x_{1t}^2 - \beta x_{2t}^2 = 0$, which defines the critical threshold ratio
		\begin{equation} \label{critical point}  
			\alpha^* = \frac{x_{10}^2 - x_{1t}^2}{x_{20}^2 - x_{2t}^2} = \frac{1 - (1 - f_1)^2}{1 - (1 - f_2)^2} \frac{x_{10}^2}{x_{20}^2}.  
		\end{equation}
		The $x_{10}:x_{20}=3:1$ rule gives the values of the critical points $\alpha^*$ to be 9.0, 9.0 and 6.12 when $(f_1,f_2)$ take the values $(0.9,0.9), (0.3,0.3)$ and $(0.3,0.5)$ respectively. Through multiple calculations with different values of $x_{10}$, it was found that the winning probability $P_\mathrm{1w}$ is independent of the numerical value of $x_{10}$. The initial state $(x_{10},x_{20})$ is simply set to $(60,20)$ in the calculations. 
		
		Under the semi-classical approximation, the dominant contribution to the path integral originates from the minimal-action path, typically obtained by solving the Euler–Lagrange equations. However, for system with Lagrangian like Eq.\eqref{nonlinear Lanchester square law}, the Euler–Lagrange equations often form a set of coupled, nonlinear differential equations that are analytically intractable. This complexity necessitates the use of numerical approaches to estimate the classical path.
		
		We first implement a Path Integral Monte Carlo (PIMC) method enhanced with simulated annealing. This algorithm discretizes the time evolution into 100 steps. The search for the dominant path proceeds through 1000 Monte Carlo cycles to encourage exploration of the configuration space. The system undergoes an annealing schedule with a cooling factor of 0.95 per cycle to progressively converge toward the solution. New path configurations are proposed via random perturbations with a step size of 0.1, while an infinitesimal offset of $1\times10^{-5}$ is applied to prevent numerical divergence. While PIMC provides a reliable benchmark, its stochastic nature prevents it from yielding an explicit functional form for the path, limiting its utility for subsequent analytical work.
		
		As a powerful alternative, we then employ the semi-analytical Rayleigh-Ritz functional optimization technique. Originally developed for Wiener path integral with constant diffusion coefficients\upcite{Petromichelakis2021}, this method is extended here to Lagrangian with a nonlinear diffusion coefficient matrix. Its implementation is described in Appendix B. 
		
\begin{figure}
			\centering
			\includegraphics[width=18cm,height=15cm]{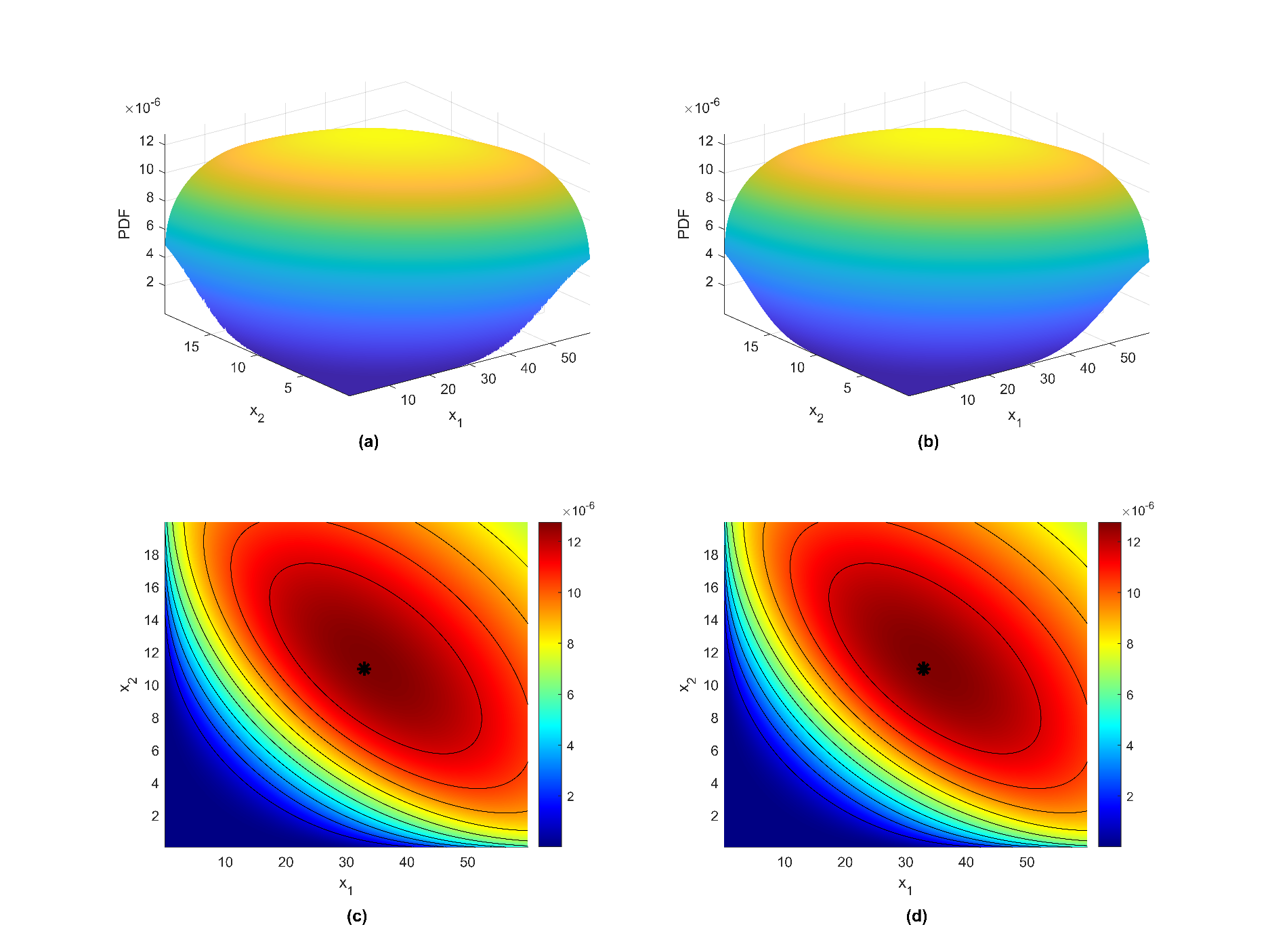}
			\caption{\label{jianyan1} (a) The PDF calculated via Rayleigh-Ritz method. (b) The PDF calculated via Path Integral Monte Carlo method. (c) Top view of (a). (d) Top view of (b). The parameters are $t=0.2,\beta=1.0,\rho=9.0$. The point marked with an asterisk (*) is the value of $(x_1,x_2)$ predicted by the Lanchester's square law \eqref{lsls}.}
		\end{figure}

		The Wiener path integral method yields the transition probability density $W(x_{1t}, x_{2t}, t|\, x_{10}, x_{20}, 0)$ as its primary outcome, expressed as a function of military strength $(x_{1t}, x_{2t})$ at time $t$. This approach fundamentally incorporates the stochastic nature of combat dynamics, accounting for fluctuations and uncertainties that deterministic models neglect. The probability density function (PDF) depicted in Figure \ref{jianyan1} is $W(x_{1t}, x_{2t}, 0.2|60, 20, 0)$ with combat efficiencies $(\rho,\beta)=(1.0,9.0)$. It illustrates a broad distribution of possible outcomes rather than a single trajectory, highlighting the probabilistic essence of the engagement. This result is distinct from the prediction of the Lanchester's square law Eq.\eqref{lsls} which yields an absolutely definite value $(x_{1t}, x_{2t})=(33,11)$ at time $t=0.2$. The Lanchester's model, being purely deterministic, does not capture the variability and random influences present in real combat scenarios. Interestingly, this deterministic value resides precisely within the region of maximum probability predicted by the Wiener path integral method, suggesting that while the Lanchester result is a likely outcome, it is by no means the only possible one. The Wiener method thus provides a more comprehensive and realistic framework by quantifying the range of potential states and their associated probabilities.
		
		The PDF in Figure \ref{jianyan1} is computed using the Rayleigh-Ritz method and PIMC, respectively. The results exhibit a high degree of consistency. This close agreement suggests both the validity and consistency of applying the Rayleigh-Ritz technique to the present problem. Furthermore, the Rayleigh-Ritz method provides a parameterized analytical expression for the minimal-action path through a well-defined convex optimization process. Thereby it can significantly improve the computational efficiency and is employed as the principal solution technique in subsequent analyses.
			
\begin{figure}
			\centering
			\includegraphics[width=18cm,height=12cm]{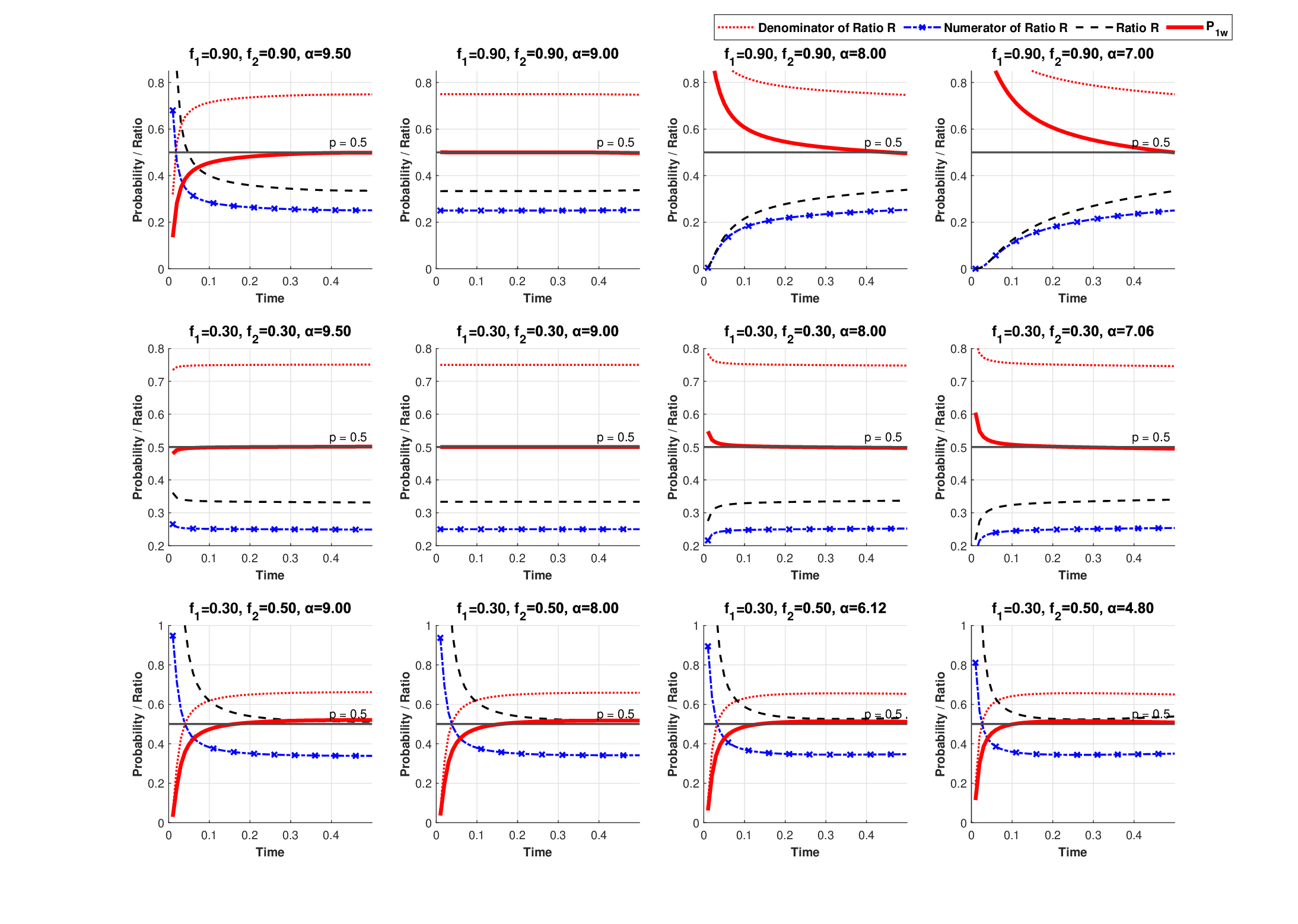}
			\caption{\label{square_law_nonlinear} 
				The temporal dynamics of $P_\mathrm{1w}$, $R$, and the constituent ratios of $R$ across discrete parameter sets $(f_1,f_2,\alpha)$.}
		\end{figure}

		The temporal dynamics of $P_\mathrm{1w}$, $R$, and the constituent ratios of $R$ across discrete parameter sets $(f_1,f_2,\alpha)$ are plotted in the Figure \ref{square_law_nonlinear}. They are shown with time plotted on the horizontal axis and the probability or ratio value on the vertical axis. The black dashed lines are the ratio $R$, and the light red dotted and light blue dash-dot curves refer to the denominators and numerators of the ratio respectively. The bold red solid curves are the normalized winning probability for the attacker $P_\mathrm{1w}$.
		
		As illustrated in Figure \ref{square_law_nonlinear}, the proposed Wiener path integral framework reveals fundamental distinctions from deterministic Lanchester predictions, primarily through the introduction of stochastic dynamics and the newly defined attacker winning probability. Within the deterministic model, force evolution and final outcome are computed using Eq.\eqref{lsls} and Eq.\eqref{sqlaw} which prescribe fixed results for any given force ratio $\alpha$. Specifically, the attacker prevails if $\alpha < \alpha^*$, fails if $\alpha > \alpha^*$, and a stalemate occurs exactly at the critical threshold $\alpha = \alpha^*$. This binary and deterministic classification leaves no room for uncertainty or probabilistic variation; each scenario leads to one and only one outcome based solely on the initial conditions. In contrast, our stochastic analysis shifts the focus from a single predetermined trajectory to a distribution of possible evolutions, thereby enabling the evaluation of the probability of attacker victory rather than a certain yes-or-no conclusion. This approach acknowledges that real-world combat scenarios are inherently influenced by random factors—such as situational uncertainties, operational discrepancies, and external disturbances—which are not captured by classical Lanchester's theory. By incorporating a Wiener process into the dynamics, the model captures the intrinsic randomness of engagement processes, reflecting more realistic conditions where outcomes are not preordained. Consequently, the stochastic framework demonstrates that, under the influence of randomness, all outcome states remain possible across the entire range of $\alpha$, though with strongly varying likelihoods. For instance, even when $\alpha$ is significantly greater than $\alpha^*$, there remains a non-zero—albeit potentially very small—probability that the attacker may still achieve victory due to favorable stochastic fluctuations. Similarly, near the critical threshold $\alpha^*$, the outcomes are particularly sensitive to random effects, resulting in a smooth transition between victory and defeat probabilities rather than an abrupt switch. This continuous and probabilistic perspective not only enriches the traditional Lanchester's model but also provides a more flexible and insightful tool for assessing combat outcomes, constituting a key innovation of our methodology.
			
		It is evident in the figure that, when combatants possess identical tolerance for troop losses, the probability of victory for the attacking force increases with decreasing $\alpha$. Conversely, when the defender maintains greater tolerance for losses, its victory probability remains largely invariant across variations in $\alpha$. For parameter configurations where $\alpha=9.0$ and $(f_1,f_2)=(0.9,0.9)$ or $(f_1,f_2)=(0.3,0.3)$, engagements evolve into stalemates under these conditions. This outcome aligns with predictions derived from the deterministic Lanchester equation. Crucially, the system exhibits distinct dynamical regimes governed by the parameter $\alpha$. In the regime $\alpha<\alpha^*$, the winning probability decays monotonically from an elevated initial value towards approximately $50\%$. This behavior indicates stable favorability towards the attacker under the 3:1 force ratio. Conversely, in the regime $\alpha>\alpha^*$, the probability ascends monotonically from a reduced baseline to approximately $50\%$, revealing the ratio's inherent disfavorability. At criticality ($\alpha=\alpha^*$), the collective fluctuation spectrum for $t > 0.5$ exhibits universality across parameter sets. This confirms its origin in intrinsic critical phenomena rather than parametric differences.	Furthermore, under asymmetric parameters $(f_1,f_2) = (0.3, 0.5)$, the system maintains defensive advantage even when $\alpha$ is less than the classical critical value $\alpha^*$, and the 3:1 force ratio remains unfavorable to the attacker. This deviation from deterministic predictions underscores the role of stochasticity and different tolerance for losses in shaping tactical outcomes.
		
		\section{Summary}
		
	In this paper, we develop a Wiener path integral formalism to incorporate stochasticity into combat dynamics based on the Lanchester's square law. This approach enables the calculation of the probability distribution governing the evolution of force strengths, offering a continuous description of stochastic engagements that goes beyond traditional deterministic formulations. Implementing this framework to analyze the empirical 3:1 combat rule, we quantitatively evaluate the temporal evolution of the attacker's victory probability across varying parameters. Our principal finding indicates that the applicability of the 3:1 rule is contingent upon specific parameters, primarily hinging on the relative combat effectiveness ratio $\alpha$ between the opposing forces and the tolerance for attrition $(f_1,f_2)$. 
	
	This study establishes a theoretical framework based on stochastic dynamics and Wiener path integral method, providing a rigorous analytical toolkit for conflict system analysis. The formalism allows for the incorporation of noise-driven dynamics and uncertainty in combat outcomes, bridging a gap between classical Lanchester's theory and real-world operational variability. Future research will extend this methodology to incorporate diverse tactical configurations of force attrition models, such as heterogeneous forces, spatially distributed engagements, and adaptive decision-making processes. Additionally, we intend to investigate significant properties, such as phase transition behaviors and correlation structures within conflict dynamics, which may offer deeper insight into the emergence of dominance and collapse in competitive systems.
		
		\section*{Acknowledgments}
		This work is supported in part by research program of NUDT.
		\section*{Appendix A: Diffusion coefficients of the Lanchester's square law}		
		\setcounterpageref{equation}{0}
		\renewcommand\theequation{A.\arabic{equation}}
		The survival status of an individual soldier in combat can be represented by a Bernoulli random variable $I$, where survival corresponds to $I=1$, and death or loss of combat capability corresponds to $I=0$. Before the battle begins, a troop has $x_{10}$ soldiers, each assigned a Bernoulli variable $I_i$ with an initial value of 1. The Bernoulli variable of a solider who is killed at time $t$ during the battle changes to 0 and remains at 0 for all future times. Then the random strength of the troop at time $t$ can be expressed by the Bernoulli random variable set \(\{I_i\}_{i=1}^{x_{10}}\)
		\begin{equation}{\label{decompose}}
			x_1(t)=I_1(t)+I_2(t)+\cdots+I_{x_{10}}(t).
		\end{equation}
		
		The infinitesimal increment form of Eq.\eqref{decompose} is
		\begin{equation}
			\mathrm{d}x_1=x_1(t+\mathrm{d}t)-x_1(t)=\mathrm{d}I_1+\mathrm{d}I_2+\cdots+\mathrm{d}I_{x_{10}},
		\end{equation}
		where $\mathrm{d}I_j$ is the infinitesimal change of $I_j$ with value $0$ or $-1$. Taking expected value $\mathrm{E}[\cdot]$ on both sides of the above equation yields
		\begin{equation}
			\mathrm{E}[\mathrm{d}x_1]=\mathrm{E}[\mathrm{d}I_1]+\mathrm{E}[\mathrm{d}I_2]+\cdots+\mathrm{E}[\mathrm{d}I_{x_{10}}]=\mathrm{d}\bar{I_1}+\mathrm{d}\bar{I_2}+\cdots+\mathrm{d}\bar{I}_{x_{10}}.
		\end{equation}
		where $\mathrm{d}\bar{I_i}=\mathrm{E}[I_i(t+\mathrm{d}t)-I_i(t)]=\mathrm{E}[I_i(t+\mathrm{d}t)]-\mathrm{E}[I_i(t)]$ is the infinitesimal variation of expected value over the interval from time $t$ to $t+\mathrm{d}t$. $\mathrm{E}[\mathrm{d}x_1]$ can be viewed as the attrition of the troop during the interval. The Lanchester's square law gives
		\begin{equation*}
			\mathrm{E}[\mathrm{d}x_1]=\sum_{i=1}^{x_{10}}\mathrm{d}\bar{I_i}=-\beta x_2 \mathrm{d}t.
		\end{equation*}
		
		Assuming that during an infinitesimal time interval $dt$, only the $j$-th soldier experiences a casualty event, while the expected values $\bar{I_i}$ of the remaining soldiers do not have time to change, it follows that for all other soldiers, $\mathrm{d}\bar{I_i}=0$, where $i\neq j$. Thus, we obtain the following expression:
		\begin{equation*}
			\mathrm{d}\bar{I_j}=-\beta x_2 \mathrm{d}t.
		\end{equation*}
		The expected value of the infinitesimal change $\mathrm{E}[\mathrm{d}I_j]$ is
		\begin{equation}{\label{probability of dI_j}}
			\mathrm{E}[\mathrm{d}I_j]=-\beta x_2\mathrm{d}t=-P(I_j(t+\mathrm{d}t)=0|I_j(t)=1),
		\end{equation}
		where $P(I_j(t+\mathrm{d}t)=0|I_j(t)=1)$ is the transition probability from state $I_j(t)=1$ to $I_j(t+\mathrm{d}t)=0$.
		The variation can be derived 
		\begin{align*}
			\mathrm{Var}[\mathrm{d}I_j] & = \mathrm{E}[\mathrm{d}I_j^2]-(\mathrm{E}[\mathrm{d}I_j])^2
			\\ &=1\cdot P(\mathrm{d}I_j=1) -(\mathrm{E}[\mathrm{d}I_j])^2
			\\& = P(I_j(t+\mathrm{d}t) = 0|I_j(t) = 1) -(\mathrm{E}[\mathrm{d}I_j])^2
			\\ &=\beta x_2 \mathrm{d}t+O(\mathrm{d}t^2).
		\end{align*}
		The assumption of the independent homogeneous distribution of $I_j$ can lead to the variation of $\mathrm{d}x_1$
		\begin{equation}\label{variation of dx}
			\mathrm{Var}[\mathrm{d}x_1]=\beta x_1x_2 \mathrm{d}t+O(\mathrm{d}t^2).
		\end{equation}
		
		From the Langevin equation \eqref{Langevin equaion} and the relationship established from Gaussian white noise $\phi_i(t)=\sigma_i\frac{\mathrm{d}W_i}{\mathrm{d}t}$\upcite{Ito1951}, we have
		\begin{equation*}
			\mathrm{d}x_1+\beta x_2\mathrm{d}t=\sigma_{11}\mathrm{d}W_1,
		\end{equation*}
		where $\mathrm{d}W$ is the infinitesimal Wiener increment. Taking the variance $\mathrm{Var}[\cdot]$ on both sides and noting $\mathrm{Var}[\beta x_2\mathrm{d}t]=O(\mathrm{d}t^2)$, along with the fact $\mathrm{Var}[\mathrm{d}W_1]=\mathrm{d}t$ from the Itô's lemma, we obtain
		\begin{equation}\label{variation of dx_1}
			\mathrm{Var}[\mathrm{d}x_1]=\sigma_{11}^2\mathrm{d}t+O(\mathrm{d}t^2).
		\end{equation}
		Combing Eq.\eqref{variation of dx} with Eq.\eqref{variation of dx_1} and ignoring the infinitesimal $O(\mathrm{d}t)$ terms, one can have
		\begin{equation}{\label{diffusion coefficient}}
			\sigma_{11}^2=\beta x_1x_2.
		\end{equation}
		With a similar derivation procedure, $\sigma_{22}^2=\rho x_1x_2$ can be derived. Therefore, for the Lanchester's square law, the diffusion coefficient matrix can be the form as Eq.\eqref{diffusion coefficient matrix}.
		
		\section*{Appendix B: The Rayleigh-Ritz functional optimization technique}		
		\setcounterpageref{equation}{0}
		\renewcommand\theequation{B.\arabic{equation}}
			The semi-analytical Rayleigh-Ritz functional optimization technique begins with determining the objective function of the optimization problem, i.e.
			\begin{equation}
				\min \mathcal{S}=\int_{0}^{\tau}\mathrm{d}t \mathcal{L} (\dot{\boldsymbol{x}}(t),\boldsymbol{x}(t),t),
			\end{equation}
			where the displacement vector is $\boldsymbol{x}=[x_1,x_2]^T$. Then, expand the path function $\boldsymbol{x}(t)$ as the approximate form
			\begin{equation}
				\boldsymbol{x}(t)=\bm{\psi}(t)+\boldsymbol{z}\boldsymbol{h}(t).
			\end{equation}
		 The $\bm{\psi}(t)$ take the l-th Hermite interpolation polynomials and its coefficient is solved to satisfy the four boundary conditions
			\begin{equation*}
				\begin{cases}\psi_1=\sum_{i=1}^{4} a_{1i}t^i
					\\\psi_2=\sum_{j=1}^{4} a_{2j}t^j
					
				\end{cases}.
			\end{equation*}
			The coefficient matrix $\boldsymbol{z}$ is a $2\times l$ matrix, which is the optimization variables in this optimization problem. Besides, $\boldsymbol{h}(t)=\left[ h_l(t)\right]_{l\times 1} $ is the test function vector which satisfies the condition $\boldsymbol{h}(0)=\boldsymbol{h}(\tau)=0$. The specific form of $h_l(t)$ is taken as the followed polynomial
			\begin{equation*}
				h_l(t)=t^2(t-\tau)^2L_l(t),
			\end{equation*}
			where $L_l(t)$ is the l-th Legendre polynomial. 
			
			Subsequently, the constrained optimization is performed with the specified parameters. This process employs a gradient-based quasi-Newton method to iteratively refine the expansion coefficients $\boldsymbol{z}$ by minimizing the discretized action functional. Starting from an initial guess, the algorithm computes the numerical gradient of the action with respect to $\boldsymbol{z}$ and updates the coefficients using BFGS-based Hessian approximations to converge toward a local minimum. Regularization is applied to prevent overfitting, while adaptive numerical integration ensures accurate gradient evaluation throughout the optimization. The entire procedure, which yields both the minimal action value and the corresponding optimal path in explicit functional form, can be conveniently implemented using MATLAB's optimization toolbox.
				
			Analogous to the proof process in the reference \upcite{Petromichelakis2021}, it follows that the Rayleigh-Ritz optimization problem for the Wiener path integral with a nonlinear diffusion coefficient matrix is a convex optimization problem. This problem necessarily possesses a global minimum, and the functional form corresponding to this optimum represents the minimal-action path functional form.
			
			The Rayleigh-Ritz method offers several distinct advantages over the PIMC approach. Most significantly, it provides a parameterized analytical expression for the minimal-action path through a well-defined convex optimization process, whereas PIMC only generates discrete path samples without explicit functional representation. This analytical foundation makes the Rayleigh-Ritz method substantially more computationally efficient, avoiding the extensive sampling required by Monte Carlo techniques. Furthermore, the method demonstrates superior numerical stability, free from the statistical fluctuations inherent in stochastic sampling approaches. The obtained functional representation enables direct theoretical analysis using standard mathematical tools and provides enhanced extensibility to more complex scenarios, advantages that the sampling-based PIMC method fundamentally lacks.

\end{document}